\documentclass[sn-mathphys,iicol]{sn-jnl}
\usepackage{caption}
\usepackage{textcomp}
\usepackage{multirow}
\usepackage{comment}
\usepackage{subfigure}
\usepackage[figuresright]{rotating}
\usepackage{threeparttable} 
\usepackage{url}
\jyear{2022}%

\begin{document}

\title[Gaseous TPC for Material Screening]{A Gaseous Time Projection Chamber with Micromegas Readout for Low Radioactive Material Screening}

\author[1]{\fnm{} \sur{Haiyan Du}}
\author[2]{\fnm{} \sur{Chengbo Du}}
\author*[1]{\fnm{Ke} \sur{Han}}\email{ke.han@sjtu.edu.cn}
\author[2]{\fnm{} \sur{Shengming He}}
\author[2]{\fnm{} \sur{Liqiang Liu}}
\author[1]{\fnm{} \sur{Yue Meng}}
\author[1,3]{\fnm{} \sur{Shaobo Wang}}
\author[1]{\fnm{} \sur{Tao Zhang}}
\author[1]{\fnm{Wenming} \sur{Zhang}}
\author[1]{\fnm{} \sur{Li Zhao}}
\author[2]{\fnm{} \sur{Jifang Zhou}}

\affil[1]{\orgdiv{INPAC; Shanghai Laboratory for Particle Physics and Cosmology; Key Laboratory for Particle Astrophysics and Cosmology (MOE), School of Physics and Astronomy}, \orgname{Shanghai Jiao Tong University}, \city{Shanghai}, \postcode{200240}, \country{China}}
\affil[2]{\orgname{Yalong River Hydropower Development Company}, \city{Chengdu}, \postcode{610051}, \country{China}}
\affil[3]{\orgdiv{SPEIT~(SJTU-ParisTech Elite Institute of Technology)}, \orgname{Shanghai Jiao Tong University}, \city{Shanghai}, \postcode{200240}, \country{China}}




\abstract{Low radioactive material screening is becoming essential for rare event search experiments, such as neutrinoless double beta decay and dark matter searches in underground laboratories.
A gaseous time projection chamber (TPC) can be used for such purposes with large active areas and high efficiency.
A gaseous TPC  with a Micromegas readout plane of approximately 20$\times$20~cm$^2$ is successfully constructed for surface alpha contamination measurements.
We have characterized the energy resolution, gain stability, and tracking capability with calibration sources.
With the unique track-related background suppression cuts of the gaseous TPC, we have established that the alpha background rate of the TPC is 0.13$\pm$0.03 $\mu$Bq/cm$^2$, comparable to the leading commercial solutions.
}


\maketitle

\section{Introduction}

For rare event searches such as neutrinoless double beta decay searches~\cite{Agostini:2017jim,0vdbd} and dark matter direct detection~\cite{Liu:2017drf,DarkMatterReview}, backgrounds from the detector itself are one of the major limiting factors for searching sensitivity. 
Low radioactive material screening is becoming a key requirement for a successful experiment. 
Radioactive contaminants in the bulk of the detector materials are of major concern, especially those that emit highly penetrating high-energy gamma rays. 
High Purity Germanium (HPGe) detectors, Inductively Coupled Plasma Mass Spectrometry (ICP-MS), and Neutron Activation Analyses are widely used to measure the bulk contamination levels of detector materials (e.g.~\cite{Abgrall:2016cct,Leonard:2017okt,PandaX-4T:2021lbm}). 
Besides the bulk contamination, radioactivities on the surface of detector parts may contribute to the region of interest of any physics search and limit the search sensitivity as well~\cite{Chen:2016qcd,  CUORE:2017ztm,PandaX:2018wtu}.
Therefore, precise determination of surface contamination is gaining more attention recently.  
Gaseous detectors have been proposed~\cite{Bunker:2013huy,Ito:2020voj,Screener3D,Pan:2022bds} for such purposes, complementing the bulk contamination detection technologies.

We have proposed the concept of Screener3D, a gaseous TPC that measures a particle's energy and tracks simultaneously with Micromegas readout modules~\cite{Giomataris:1995fq}.
For a large readout plane of 2000~cm$^2$ and background suppression with tracks, we estimated the sensitivity of Screener3D is about 100~$\mu$Bq/m$^{2}$ of a 2-days run for typical measurements.
In this paper, we describe the construction of a prototype TPC of Screener3D, as well as the detector performance and background rate of the detector.

\section{Detector design and performance}

\subsection{Key detector components}
\begin{figure}
\centering
\includegraphics[width=\columnwidth]{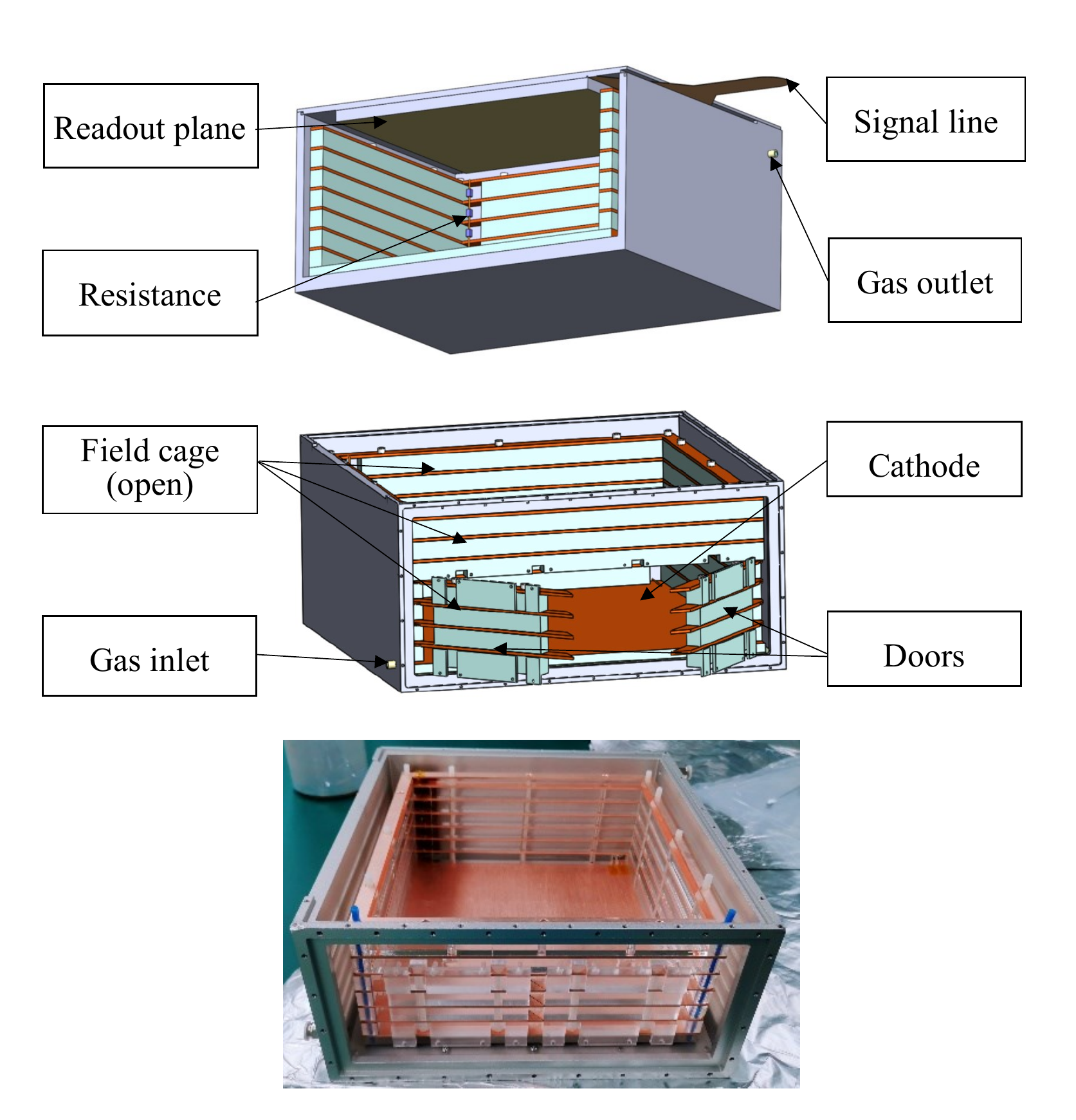}
\caption{Schematic drawing of the gaseous detector with key components highlighted. 
	Micromegas is attached to the top with the signal tail extended outside the aluminum vessel.
}
\label{fig:grabTPC}
\end{figure}

As shown in Fig.~\ref{fig:grabTPC}, the prototype TPC consists of an aluminum vessel, a field cage, a cathode, and a Micromegas readout plane.
The field cage encloses a volume of 20$\times$20$\times$9.8~cm$^3$ and is made of acrylic and copper bars stacked together alternately. 
The acrylic bars are 1.2~cm thick and copper bars 0.2~cm thick.
The front wall of the field cage can be opened as two doors for placing samples.
The vertically neighboring copper bars are connected by 1~G$\Omega$ resistors working as a voltage divider to form uniform drifting electric fields in the field cage.
A 2~mm thick polished copper plate is attached to the bottom of the field cage as the cathode, which is supplied a high voltage while the TPC is operating.
The overall assembly follow resembles that field cage design in~\cite{Lin:2018mpd}.
On the top of the field cage, a faced-down 20$\times$20~cm$^2$  Micromegas module serves as the readout plane.
The Micromegas is fabricated with the thermal bonding technique~\cite{TBMM}.
A total of 128 strips are read out with 64 channels in the X and Y direction each.
A strip is made of interconnected diamond shapes with a pitch distance of 3~mm, the same design as the Micromegas described in~\cite{Lin:2018mpd}.
The Micromegas is fixed to an aluminum backplane and then screwed to the top of the aluminum vessel.
The aluminum vessel can be opened from the top and the front to facilitate the installation of the Micromegas and loading of the samples.
The strip signal and mesh high voltage line of the Micromegas are on a long tail made of 200~$\mu$m-thick Kapton-based printed circuit.  
The tail is sandwiched between the top panel and the side of the aluminum vessel.
The signals can be conveniently read out and mesh voltage applied from the outside of the vessel. 
The air-tightness of the vessel is not affected significantly because of the thinness of the tail and with the help of rubber O-rings for sealing.
Before assembly, all the detector components except the Micromegas are cleaned following the procedures outlined in ~\cite{ortec_Mega_Fu, QZC} to remove the surface contaminations such as $^{210}$Po, $^{232}$Th, and $^{238}$U.

\begin{figure}[tb]
\centering
\includegraphics[width=\columnwidth]{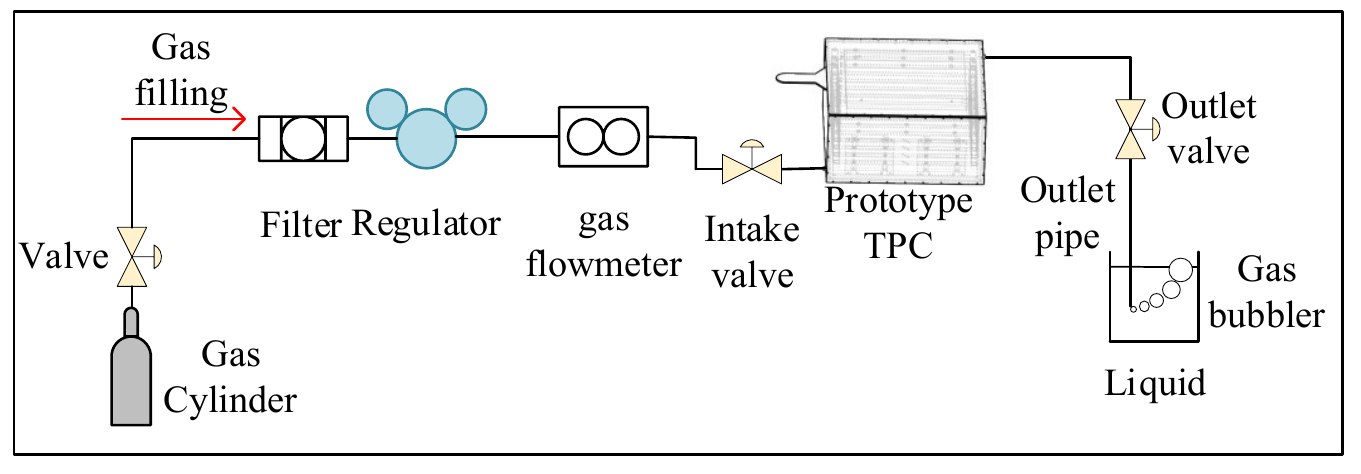}
\caption{Schematic diagram of the prototype TPC gas system.}
\label{fig:gasSystem}
\end{figure}

The TPC is operated in the flow-gas mode with Ar-CO$_2$ (or Ar-isobutane) mixture gas continuously flushed in and out of the vessel at fixed flow rates.
Ar-CO$_2$ mixture is selected as the working medium since it is widely available commercially and non-flammable. 
The gas system is shown in Fig.~\ref{fig:gasSystem}.
The gas mixture first passes through a particulate filter to prevent possible dust from entering the detector. 
A mass flow controller after the regulator provides a stable and controllable flow rate for flushing the TPC.
To prevent air from flowing back into the TPC, the gas outlet pipe is connected to a bubbler.

A NIM-based high voltage module (Iseg NHR 42 20r) provides negative bias voltages for cathode and Micromegas.
The electronics~\cite{electronics} used for the readout of Micromegas are based on the AGET chip, which can sample at frequencies ranging from 1~MHz to 100~MHz~\cite{AGET}. 
Each channel of the AGET chip integrates a charge--sensitive amplifier(CSA), which has four dynamic ranges (120~fC, 240~fC, 1~pC, and 10~pC), an analog filter, a discriminator, and 512-sample memory.
For the signals readout, we use the multiplicity mode, in which the AGET chip compares the input of each channel to a pre-defined threshold.
Once signals from multiple channels exceed the threshold, the event is registered, and then all the pulses over the threshold are recorded on disk.

\subsection{Optimized run conditions}

The run conditions are optimized for detector energy resolution and long-term stability. 

\begin{figure}[tb]
\centering
\includegraphics[width=\columnwidth]{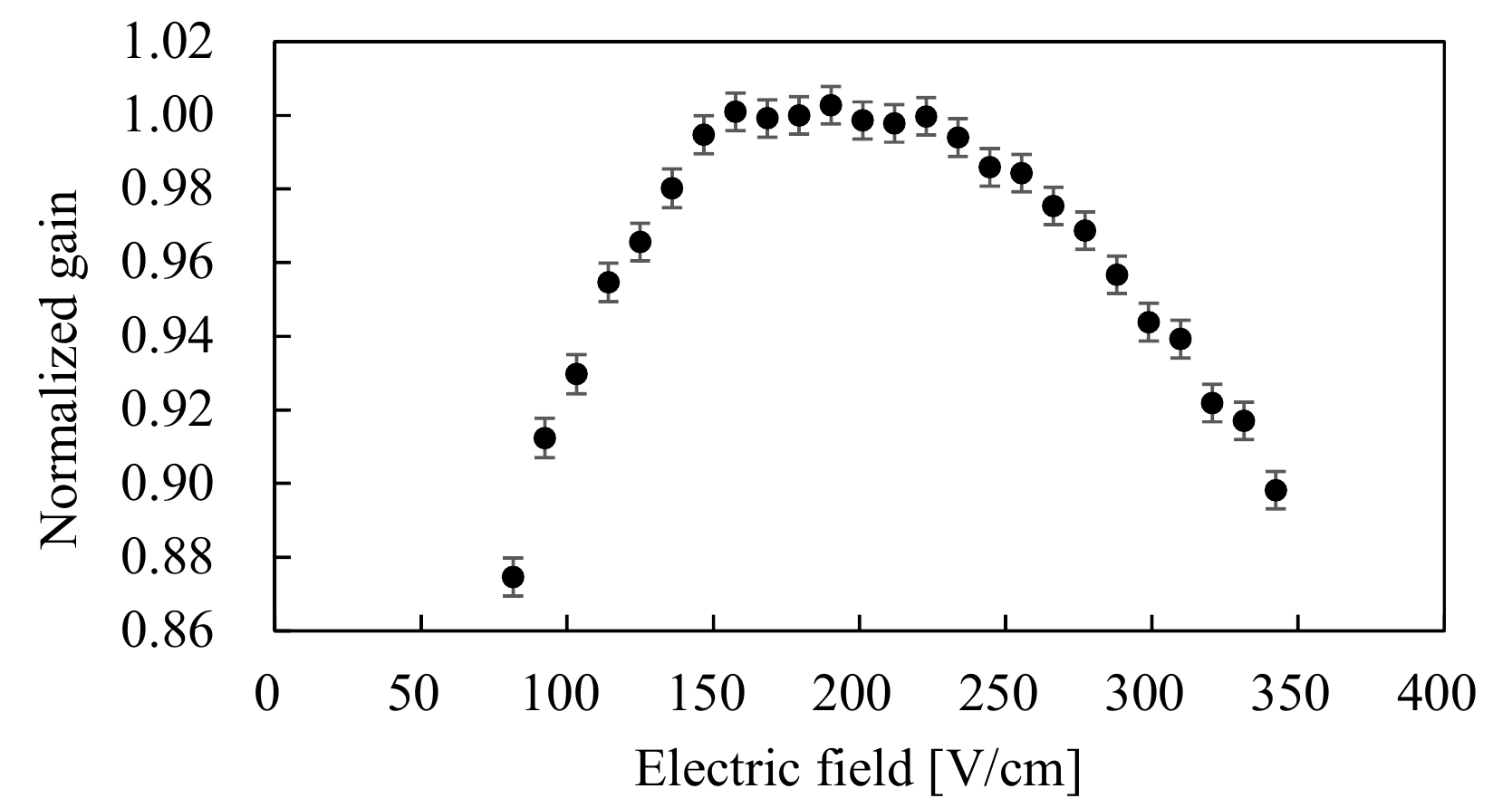}
\caption{The normalized gain as a function of the drifting electric field.
The Micromegas avalanche amplification field is set to 3500~V/cm.}
\label{fig:GainVsElectriFiled}
\end{figure}

For the 1 bar Ar-7\%CO$_2$ gas mixture, the optimal avalanche electric field of Micromegas is pre-determined based on dedicated tests and set to be 3500~V/cm. 
The TPC drift field is then varied to characterize its correlation and detector gain, as shown in Fig.~\ref{fig:GainVsElectriFiled}.
As the drifting electric field increases, the TPC gain increases because electrons and ions are less likely to recombine.
The gain plateaus when the recombination of electron and ion is negligible.
After that, the gain decreases due to the mismatch between the drift field and the Micromegas avalanche electric field.
The mismatch prevents drift electrons from entering the Micromegas avalanche region effectively.
During our operation with 1 bar Ar-7\%CO$_2$ gas mixture, the drifting field is set at 179~V/cm, which corresponds to the center point of the gain plateau.

\begin{figure}[tbp]
\centering
\includegraphics[width=\columnwidth]{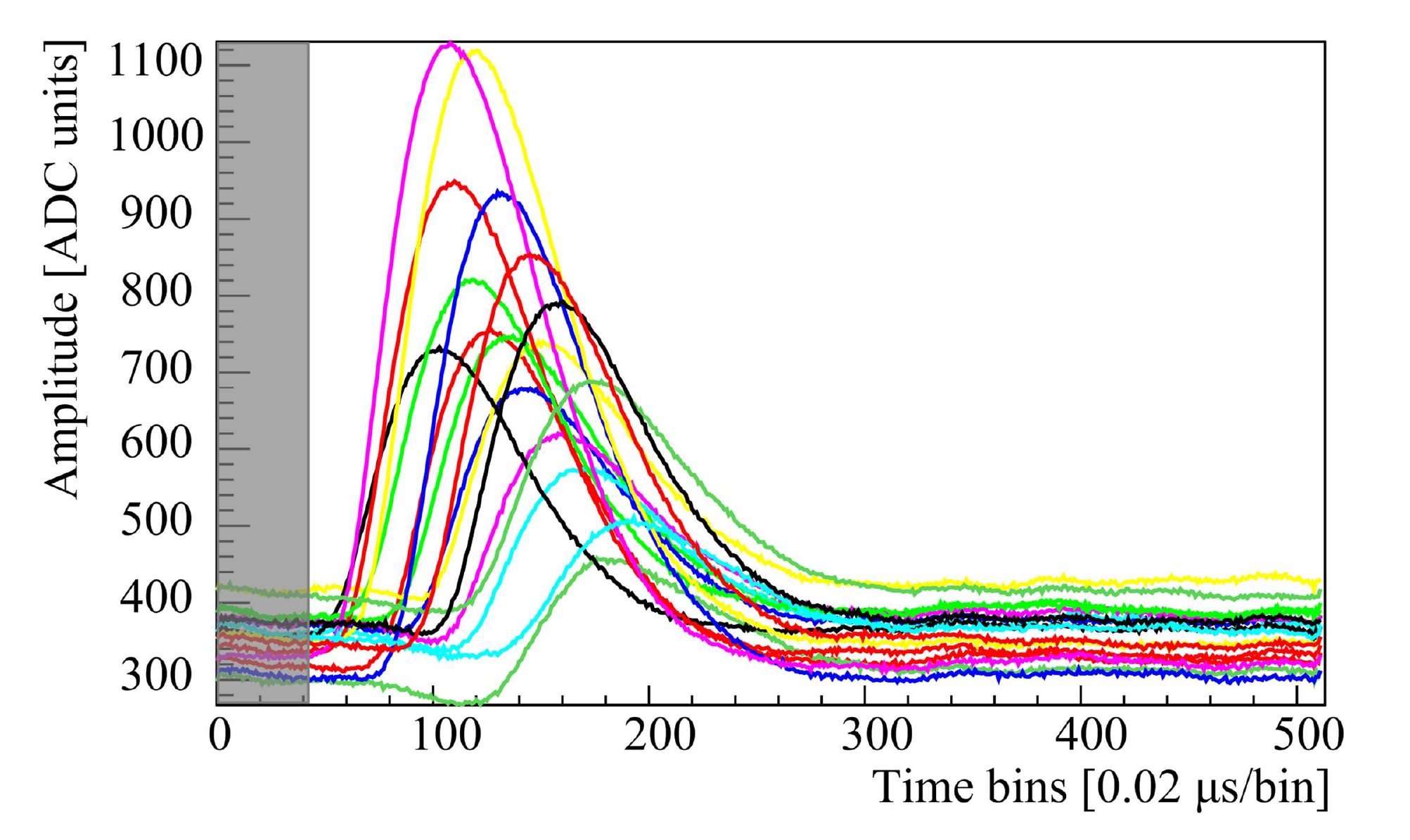}

\caption{Pulse signals of an alpha event. 
Each pulse represents a triggered channel of Micromegas.
The first 40 sampling points (shaded) are used to calculate baselines of each channel.}
\label{fig:Am241_pulse_tack}
\end{figure}

The configurations of readout electronics are optimized for alpha detection.
Based on the Garfield simulation, the electron drift velocity is 1.76~cm/$\mu$s resulting in a drifting time of 5.57~$\mu$s for the 9.8~cm drift length~\cite{Veenhof:1993hz,Garfield}.
We use a sampling frequency of 50 MHz, resulting in a time window of 10.24~$\mu$s for 512 samples to cover the full drift length.
The dynamic range of CSA is selected to be 1 pC, resulting in less than 2\% of signals from 5.5~MeV $\alpha$ particles being saturated in our measurement.
The trigger threshold is approximately 0.024 pC above the baseline.
The pulses of an alpha event from $^{241}$Am are shown in Fig.~\ref{fig:Am241_pulse_tack}.
The high threshold prevents electrons, gamma rays, or muons from triggering.

\subsection{Energy calibration and gain stabilities}

\begin{figure}[tbp]
	\centering
	\subfigure[Data in Ar-7\% CO$_2$.]
	{\includegraphics[width=0.70\columnwidth]{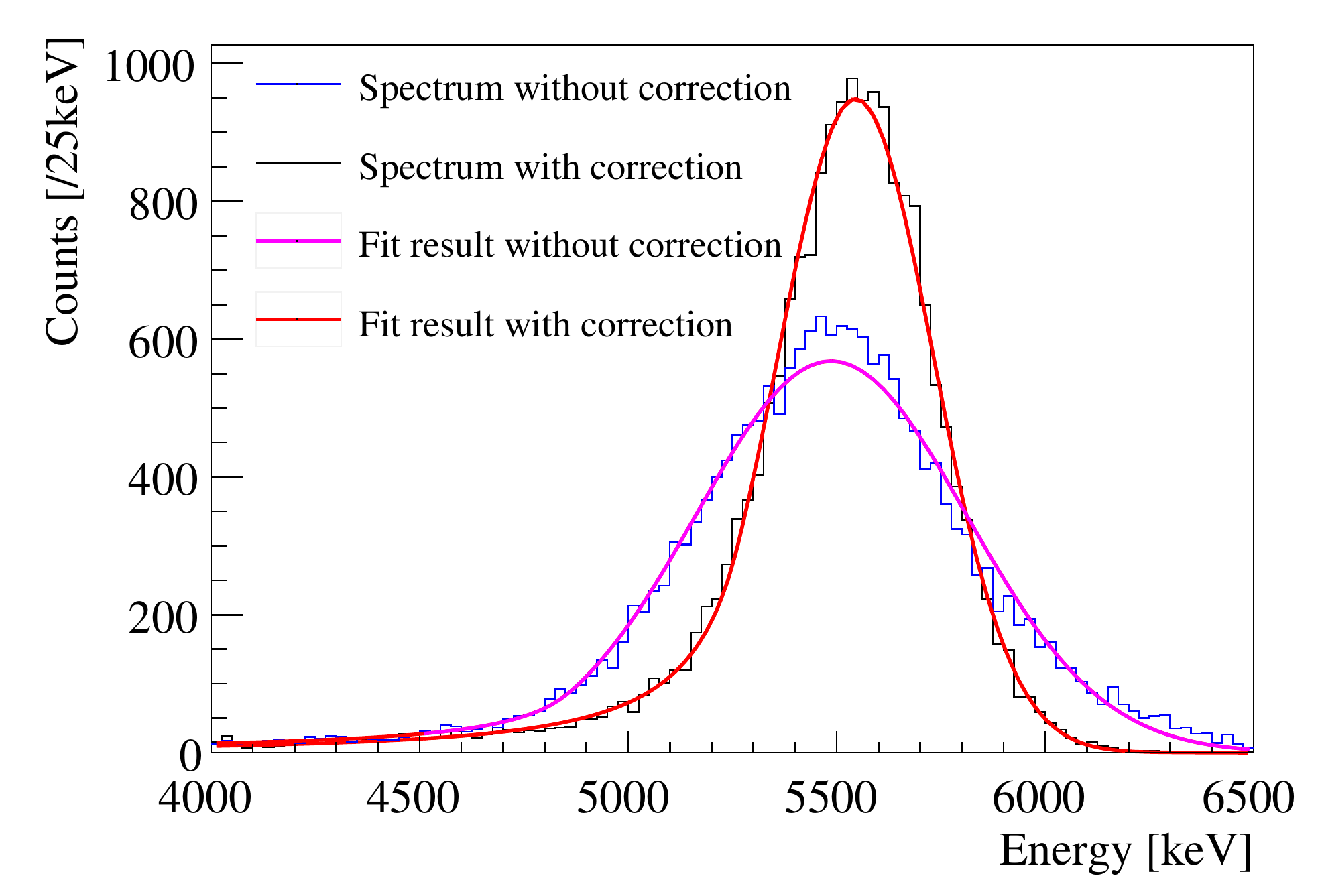}
	\label{fig:EresoCO2}}
	\subfigure[ Map of the gain correction factor.]
	{\includegraphics[width=0.70\columnwidth]{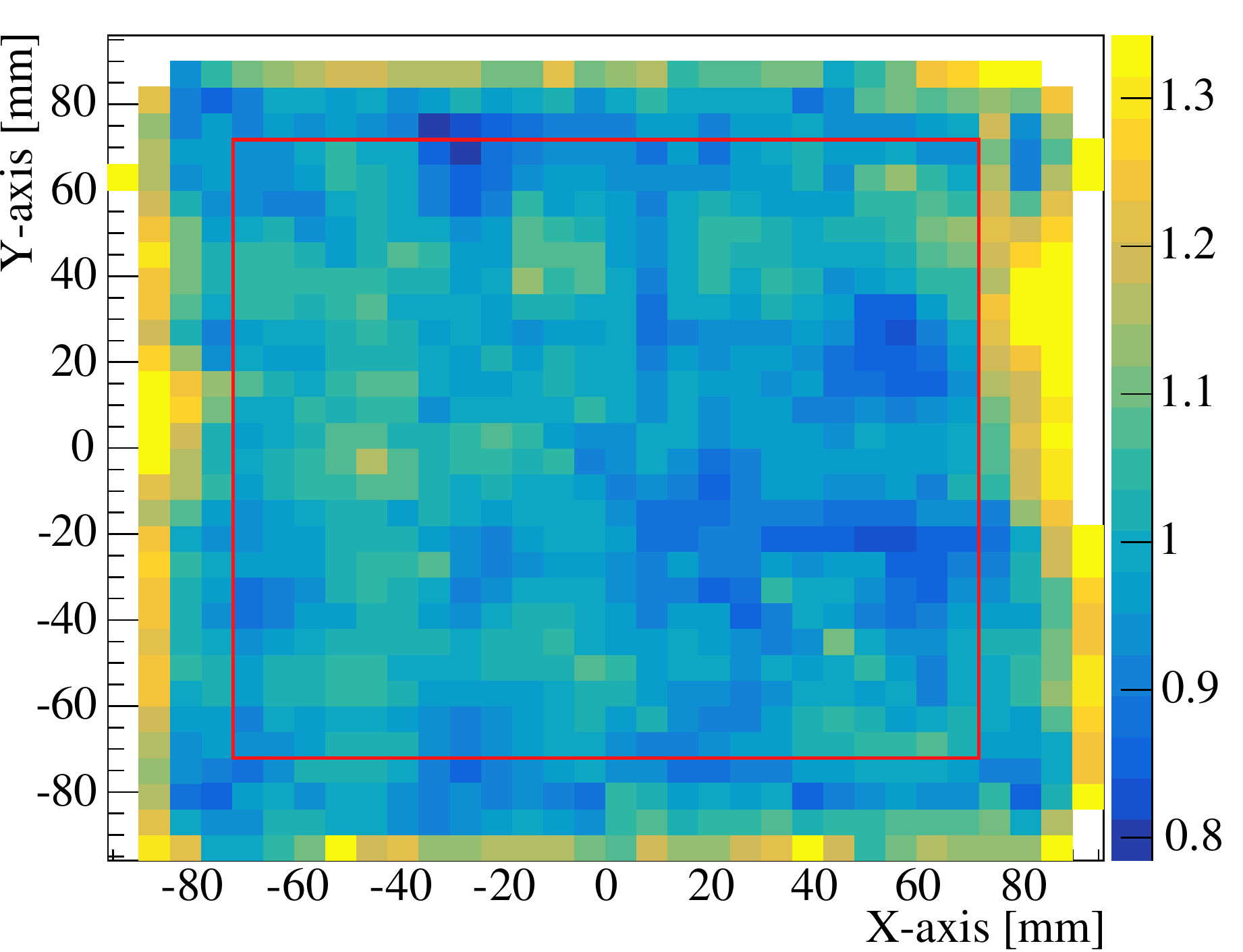}
	\label{fig:mapping}}
	\subfigure[Data in Ar-2.5\% isobutane.]
	{\includegraphics[width=0.75\columnwidth]{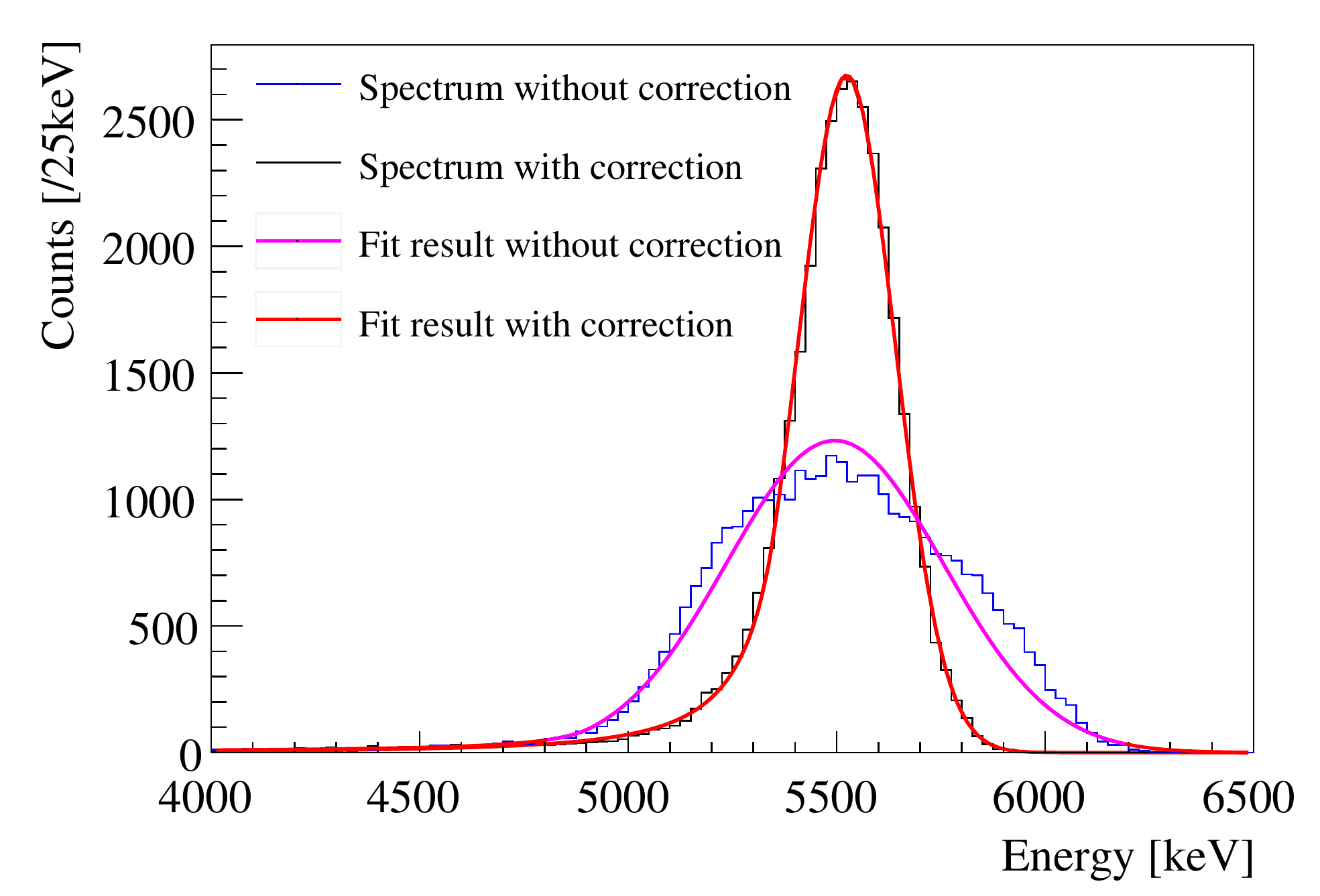}
	\label{fig:EresoISO}}
	
	\caption{(a) The energy spectrum of the alpha events from the $^{241}$Am source in Ar-7\%CO$_2$ gas mixture before and after gain correction. (b) Map of the relative gain correction in different parts of the Micromegas. The red square denotes the fiducial volume (discussed later).(c) The same as (a) but with Ar-2.5\%isobutane gas.}
	\label{fig:Am241_EnergySpec}
	\end{figure}

The detector performance is calibrated with an $^{241}$Am alpha source of 6~mm in diameter.
The energy of each event is calculated by summing up charges collected by each triggered strip.
For each strip, the baseline value is calculated by averaging the data collected in the first 40 sampling points of the pulse, as shown in Fig.~\ref{fig:Am241_pulse_tack}.
The charge collected on each strip is the integral of all sampling points minus the baseline value.
A measured energy spectrum of $^{241}$Am is shown in Fig.~\ref{fig:EresoCO2}. 
The peak corresponds to the 5.4~MeV alpha peak from the source. 
The spectrum is fitted with the crystal ball function~\cite{Oreglia:1980cs} and the obtained energy resolution is 13.6\% (Full-Width at Half Maximum) in Ar-7\%CO$_2$ gas mixture.

The energy resolution can be improved by correcting inhomogeneous gains of different positions of the Micromegas.
The readout plane is evenly divided into 32$\times$32 regions.
The $^{241}$Am source is placed at different positions to collect data covering the entire readout plane. 
The mean response in different regions is regarded as the gain correction factor of inhomogeneity, as shown in Fig.~\ref{fig:mapping}.
The maximum correction factor by the edge of the Micromegas is about 1.35, and the majority of the correction factors in the fiducial volume (discussed later) are in the range of 0.9 and 1.1.
The correction factors and the energy spectrum to be corrected are not from the same data but are collected under identical running conditions, including the gas mixture.
As one can see in Fig.~\ref{fig:Am241_EnergySpec}, the energy spectrum of $^{241}$Am is improved significantly to 9.5\% FWHM with gain corrections.
Similar improvement is also observed with a one-bar gas mixture of Ar and 2.5\% isobutane, as shown in Fig.~\ref{fig:EresoISO}.
The energy resolution before and after gain correction is 12.1\% and 4.8\% FWHM respectively.

\begin{figure}[tb]
\centering
\includegraphics[width=0.9\columnwidth]{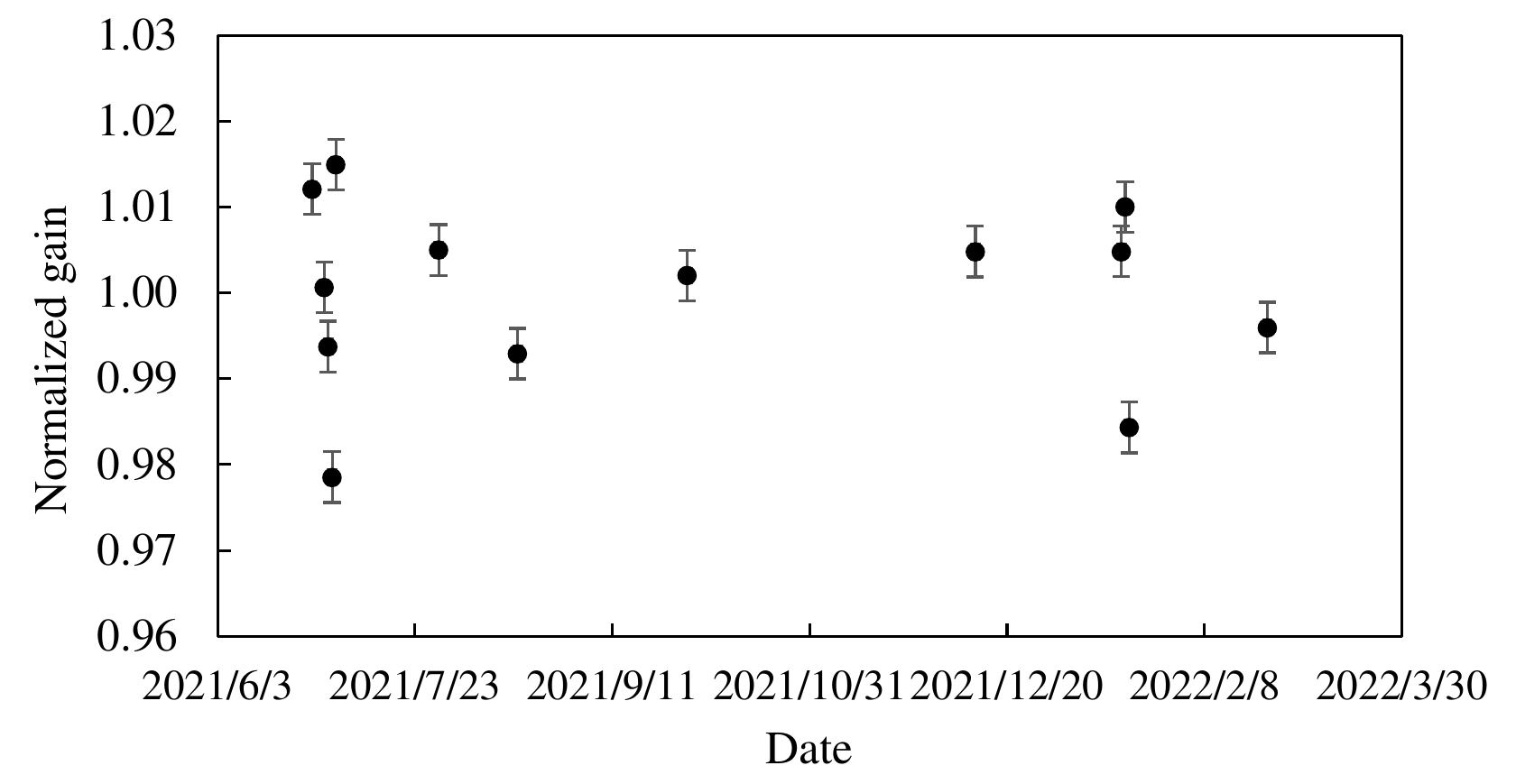}
\caption{The normalized gains measured during the 8-months monitoring.}
\label{fig:stability}
\end{figure}

We constantly monitor the gain stability of the TPC with the $^{241}$Am source.
In Fig.~\ref{fig:stability}, the normalized gain is shown and the deviation of the gain is relatively stable within 3.7\% for data taken over 8 months of operation.
It is worth noting that the detector was not turned on or flushed with a gas mixture most of the time.
The gain stability is of vital importance for our measurements over an extended period.

\subsection{Counting efficiency}
\begin{figure}[tb]
	\centering
	\includegraphics[width=0.8\columnwidth]{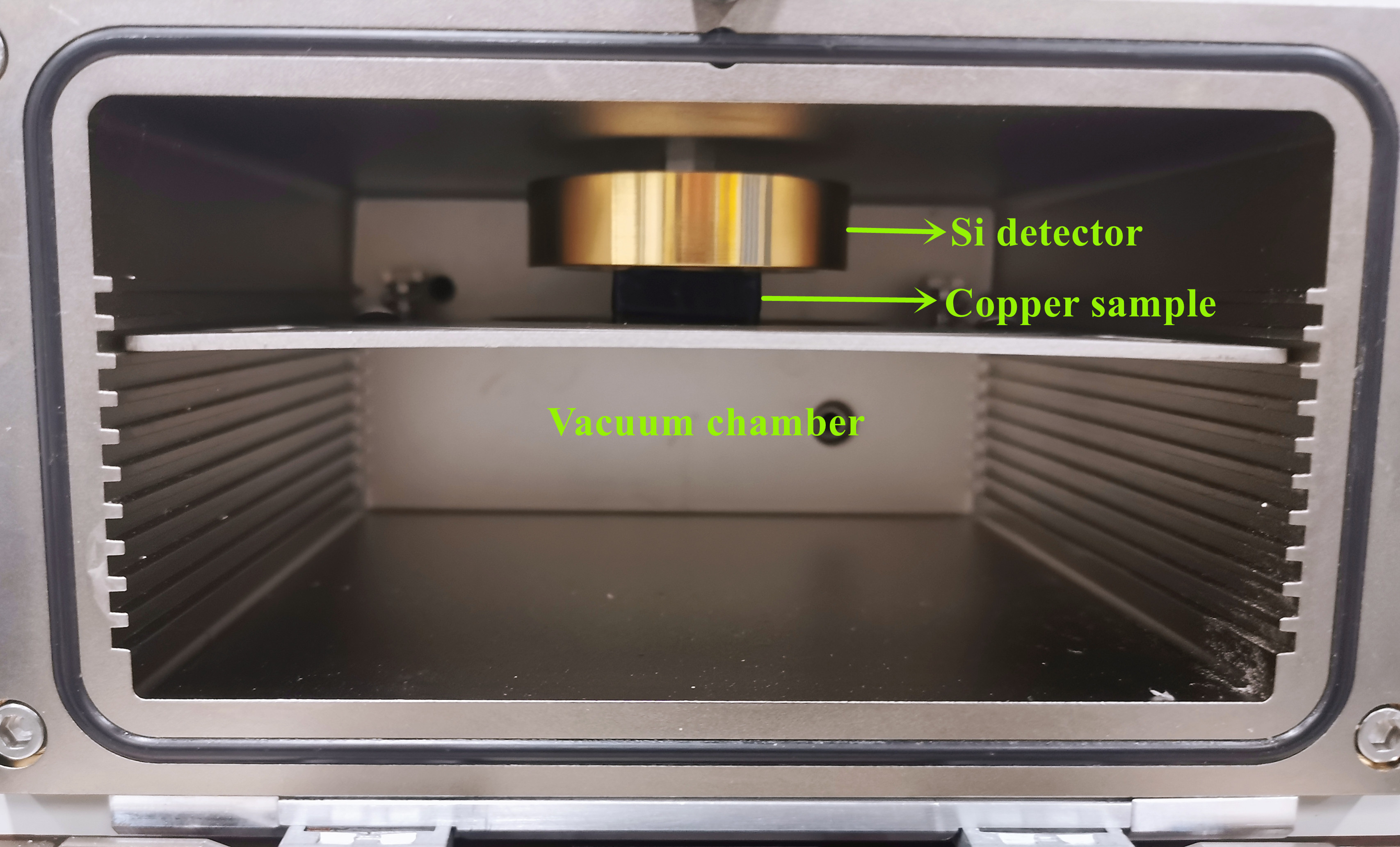}
	\caption{The copper sample is placed directly facing the detector when measured with the Ortec Alpha Mega detector.
	The chamber is pumped to a rough vacuum to minimize energy loss in air.}
	\label{fig:ortec_Mega}
	\end{figure}

	\begin{table}[tb]
	\centering
	\caption{Count rates of the alpha events measured by the prototype TPC and Ortec Alpha Mega detector. }
	\begin{tabular*}{\hsize}{@{}@{\extracolsep{\fill}}cccc@{}}
		\hline
		Source      &Detector        &Counts/s                             &Counts/s         \\
		&                &                                     &after corrections \\ 
		\hline
		$^{241}$Am  &TPC   &364.8$\pm$3.5                        &438.5$\pm$15.4   \\		
		$^{241}$Am	 &Ortec     &376.8$\pm$1.1                        &416.8$\pm$8.4   \\
		Copper	     &TPC  &(14.3$\pm$0.3)$\times$10$^{-3}$      &(14.4$\pm$0.3)$\times$10$^{-3}$    \\	
		Copper      &Oretc      &(13.2$\pm$0.5)$\times$10$^{-3}$     &(13.9$\pm$0.6)$\times$10$^{-3}$       \\			
		\hline
	\end{tabular*}
	\label{tab:Am_copper}
	\end{table}
	
The counting efficiency is calculated based on simulation and cross-validated with a commercial silicon detector. 
A dedicated simulation program based on GEANT4 has been developed to study the detector response~\cite{Screener3D,Altenmuller:2021slh}.
In addition to the $^{241}$Am calibration source, we also measured a copper sample intentionally contaminated with alpha radioactivities.
The copper sample has a surface area of 1.9$\times$0.9~cm$^2$ and has been exposed to $^{226}$Ra.
The main alpha emitter is the progeny $^{210}$Po attached to the surface of the sample.
When both are placed in the center position of the detector, the measured rates are $364.8\pm3.5$ and $(14.3\pm0.3)\times10^{-3}$ counts/s respectively.
The simulated detection efficiency is $99.0\pm0.3\%$ ($99.6\pm0.3\%$) for the calibration source (copper sample).
Besides the efficiency, a correction factor for dead time is $1.19\pm0.04$ is applied for the $^{241}$Am alpha source measurement due to the high counting rate. 
The dead time correction is negligible for the copper sample.
The final event rates are listed in Table~\ref{tab:Am_copper}.

Both the $^{241}$Am alphas source and the copper sample are measured with an Ortec Alpha Mega detector~\cite{OrtecMega}.
The Alpha Mega detector has a relatively large active area of 12~cm$^2$.
The samples can be placed directly facing the detector. 
An independent Monte Carlo simulation shows the counting efficiencies are $90.4\pm1.8\%$ and $94.8\pm1.5\%$ respectively for $^{241}$Am and copper.
The efficiency loss is mainly from the geometrical effect of the solid angle coverage. 
The measured rates and efficiency corrected rates are also listed in Table~\ref{tab:Am_copper}.
It should be noted that Alpha Mega applies dead time correction automatically when reporting event rates. 

The results from the two detectors are consistent statistically and confirm the counting efficiency calculation of our TPC.
The cross-validation also assures that there is no significant efficiency loss in our data taking or data analysis chain.

\section{Background suppression with tracks}

\begin{figure}
	\centering
	\includegraphics[width=\columnwidth]{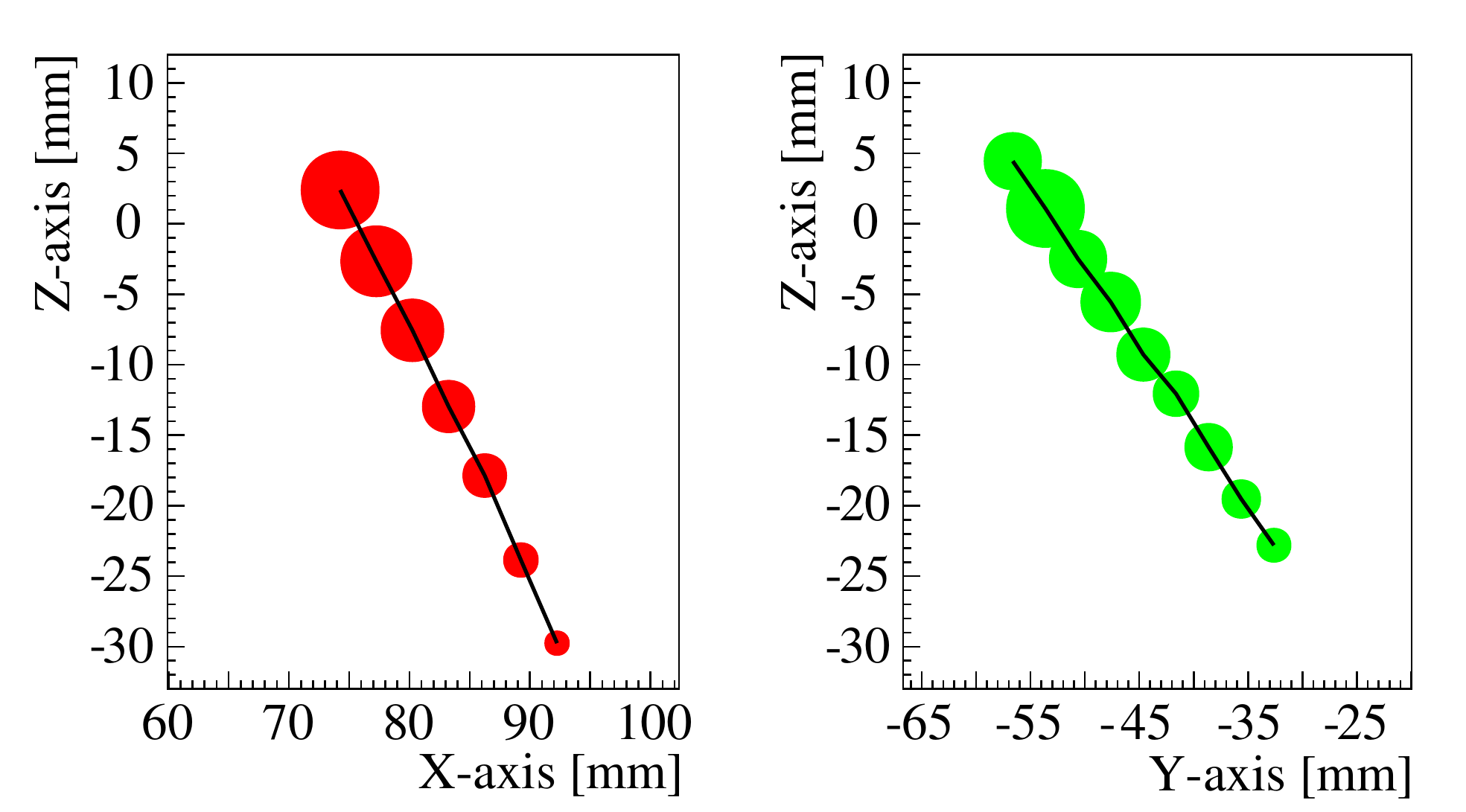}
	\caption{Reconstructed tracks of a typical alpha event in X-Z and Y-Z plane respectively. Each red/green dot means a triggered strip. The size of the dots represents the amount of deposited energy in the strip. The black line is the reconstructed track by connecting nearby hits.}
	\label{fig:tracks}
\end{figure}

\begin{figure}[tbp]
	\centering
	\subfigure[Distribution of reconstructed starting points.]
	{\includegraphics[width=0.75\columnwidth]{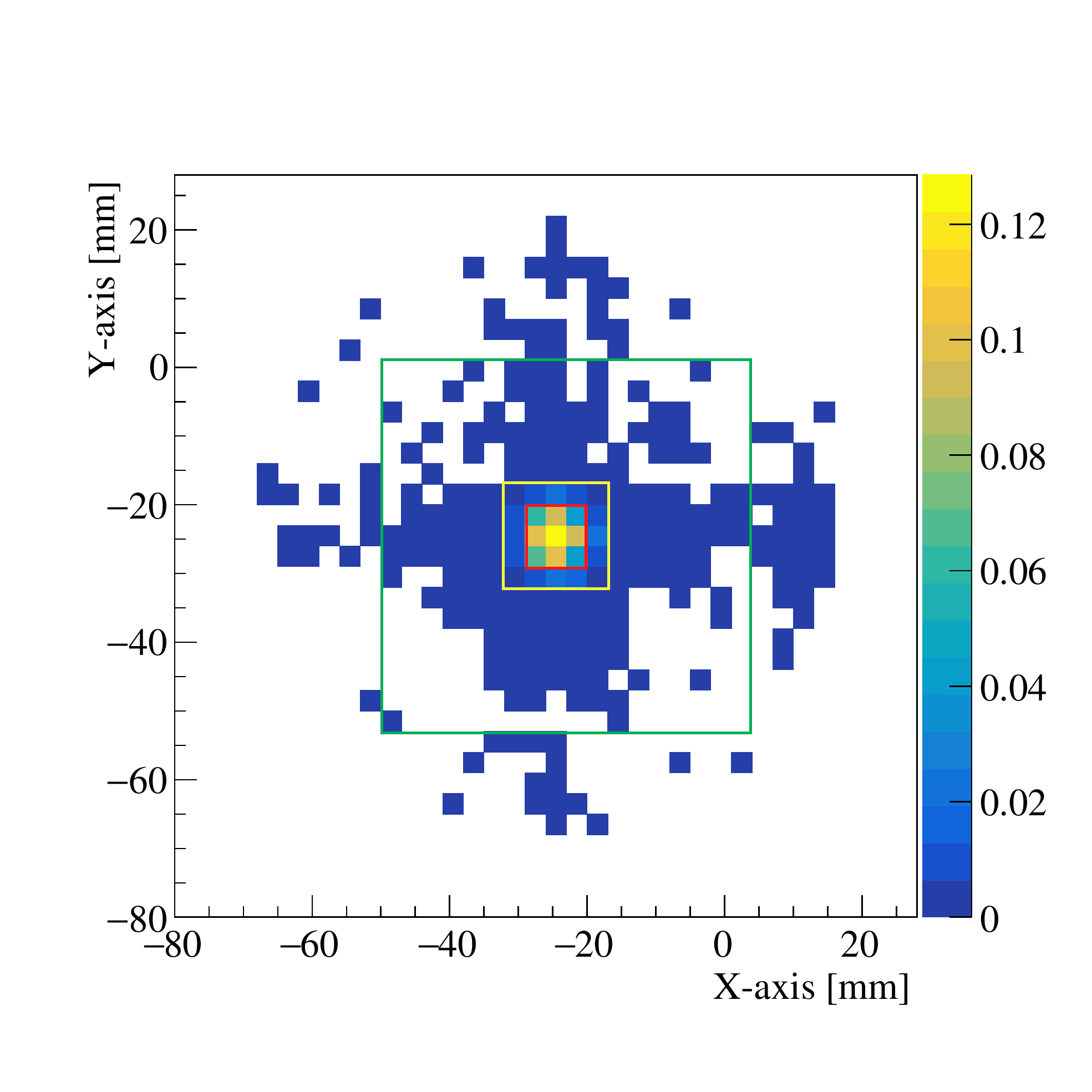}
		\label{fig:origin}
	}	
	\subfigure[Distribution of reconstruction track orientations.]
	{\includegraphics[width=0.75\columnwidth]{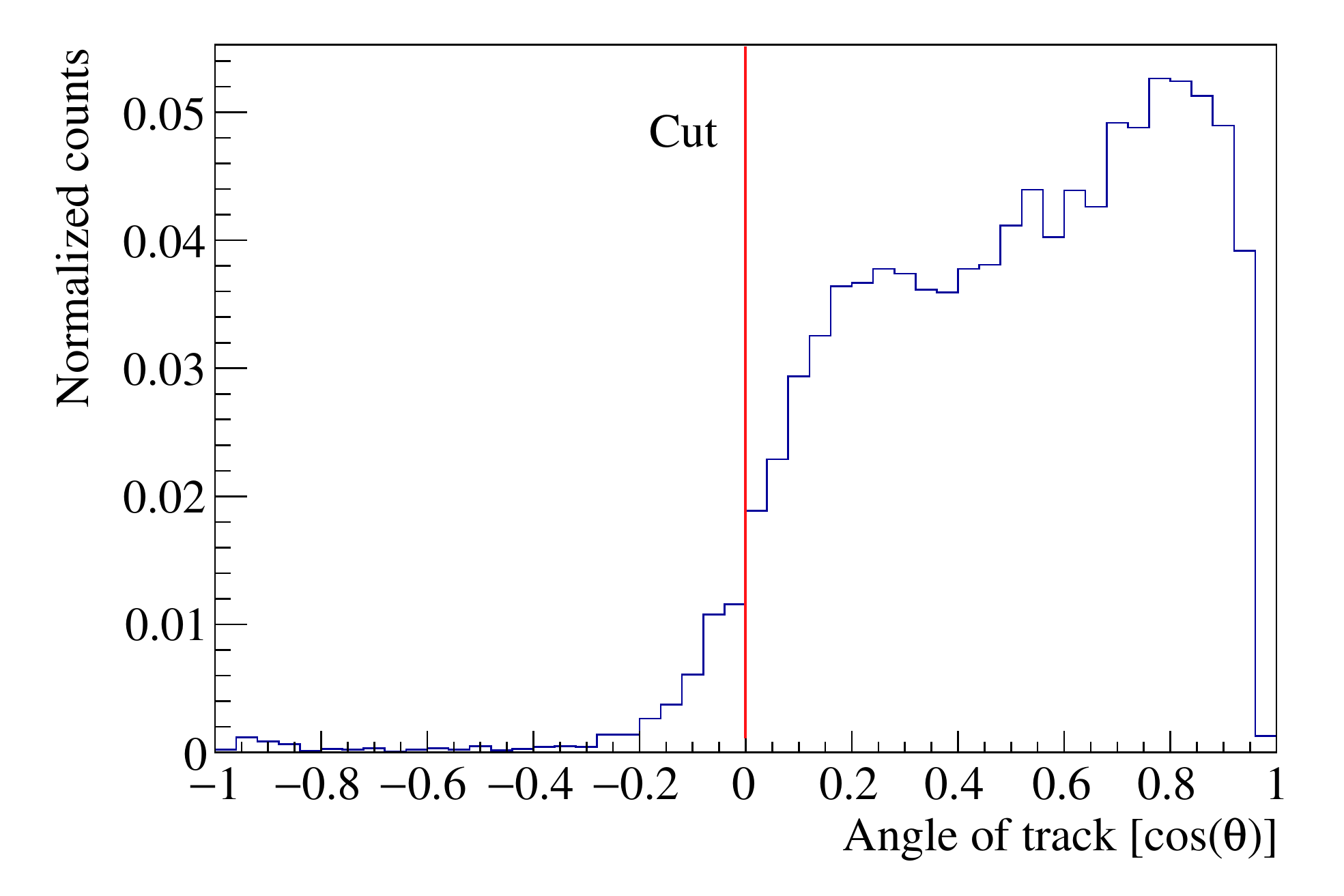}
		\label{fig:source_angle}}	
	\subfigure[Distribution of nHits.]
	{\includegraphics[width=0.75\columnwidth]{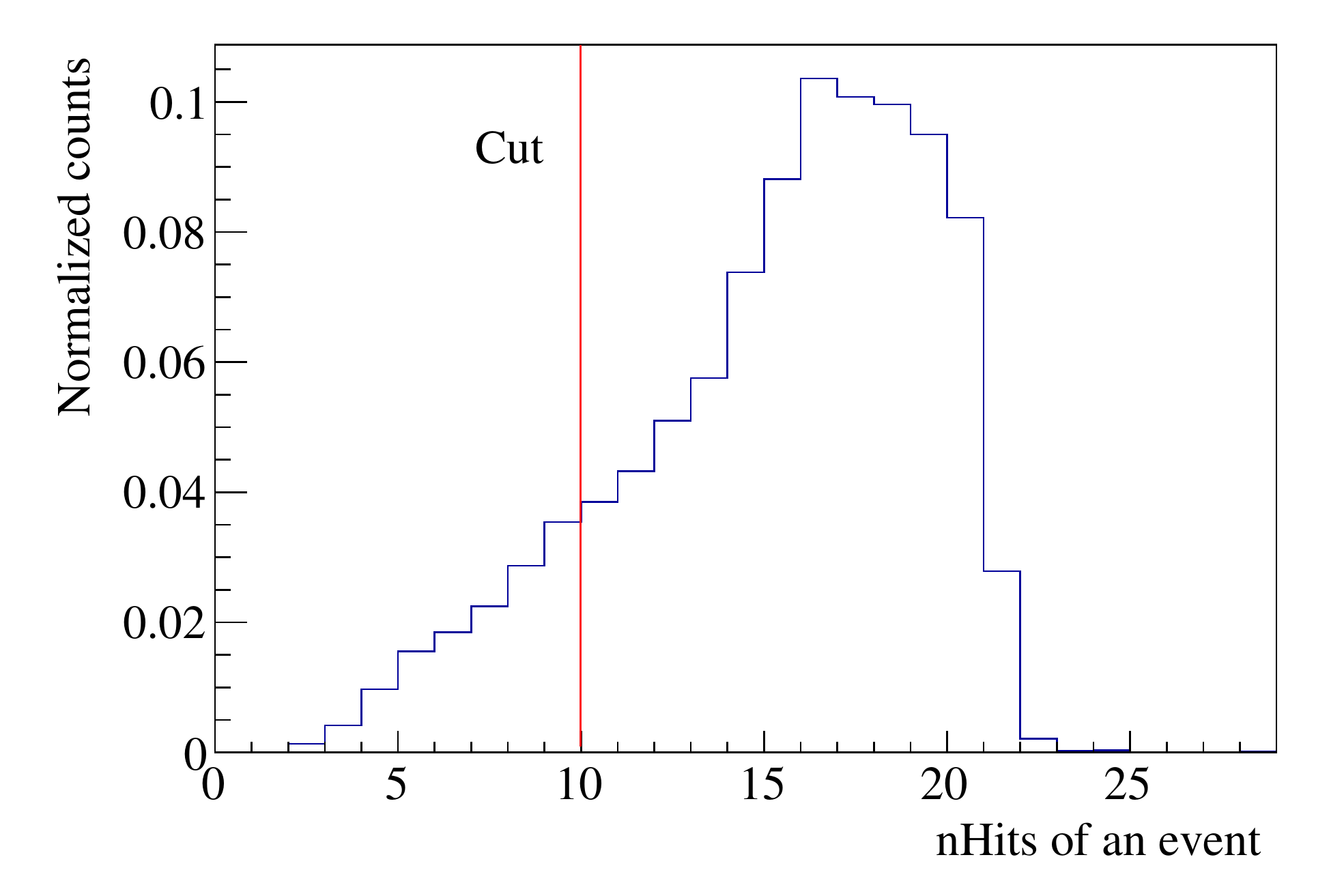}
		\label{fig:source_nHits}}	
	\caption{Distribution of the track-related parameters from alpha events of the $^{241}$Am source. 
		Please see the text for the definition of squares in (a).
	}
	\end{figure}
	
The intrinsic background of the TPC itself limits the sensitivity for low-radioactive material screening.
Gaseous TPC measures not only the energy but also the track of an alpha event in the gas medium.
We can utilize the starting position, the track angle, and the number of triggered strips (called hits for short) to select alpha events emitted from a measurement sample.

In 1 bar argon gas, the high-energy alpha particles travel in a relatively straight trajectory and reconstruction of the track is straightforward in most cases.
Fig.~\ref{fig:tracks} demonstrates one example of reconstructed tracks in the X-Z and Y-Z planes, where Z denotes the vertical drift direction and X/Y denotes the horizontal plane. 
Z is calculated based on the relative timing among the pulse on the strips and drift velocity.  
It should be pointed out that Z represents the relative not absolute Z position, but the ambiguity introduces marginal impact to alpha track reconstruction.
Along the straight line in the two-dimensional tracks, one can identify the energy deposited along the track, and the difference in each strip is significant.
The energy deposition difference originates from the Bragg effect and can be used to identify the starting point of the track, where $dE/dx$ is smaller. 
Combing the two-dimensional X-Z and Y-Z tracks, we can also calculate the orientation of the track with respect to the vertical direction.  
We also count the number of triggered strips in X and Y directions, called nHits, as an indicator of measured energy. 

Tracks from alpha particles emitted from a sample placed on the cathode have distinctive features that can be used to enhance signal identification and background rejection efficiencies.
The tracks should all be high energy, upward-going, and with a starting point within the area of the sample.
We can reject the backgrounds of the field cage and cathode with the starting points as they are distributed at the edge of the sensitive volume.
Alpha tracks from the Micromegas are all directed downward and can be rejected.
Alpha particles emitted from the impurities in the Ar-CO$_2$ mixture travel in the TPC isotropically and a track angle cut rejects 50\% of the events ideally. 
Furthermore, high-energy alphas originating from the gas medium itself have a relatively large opportunity to escape the active volume without depositing the full energy in the gas. 
Therefore, those events tend to have short tracks.
Similar to what we have reported in~\cite{Screener3D}, we have used a fiducial cut with a fiducial distance from the inner surface of the field cage larger than 27~mm, an angle cut to select upward-going tracks, and a nHits$\geq$10  cut to suppress background.

\begin{figure}[tb]
	\centering
	\subfigure[Experimental data]{
			\includegraphics[width=0.75\columnwidth]{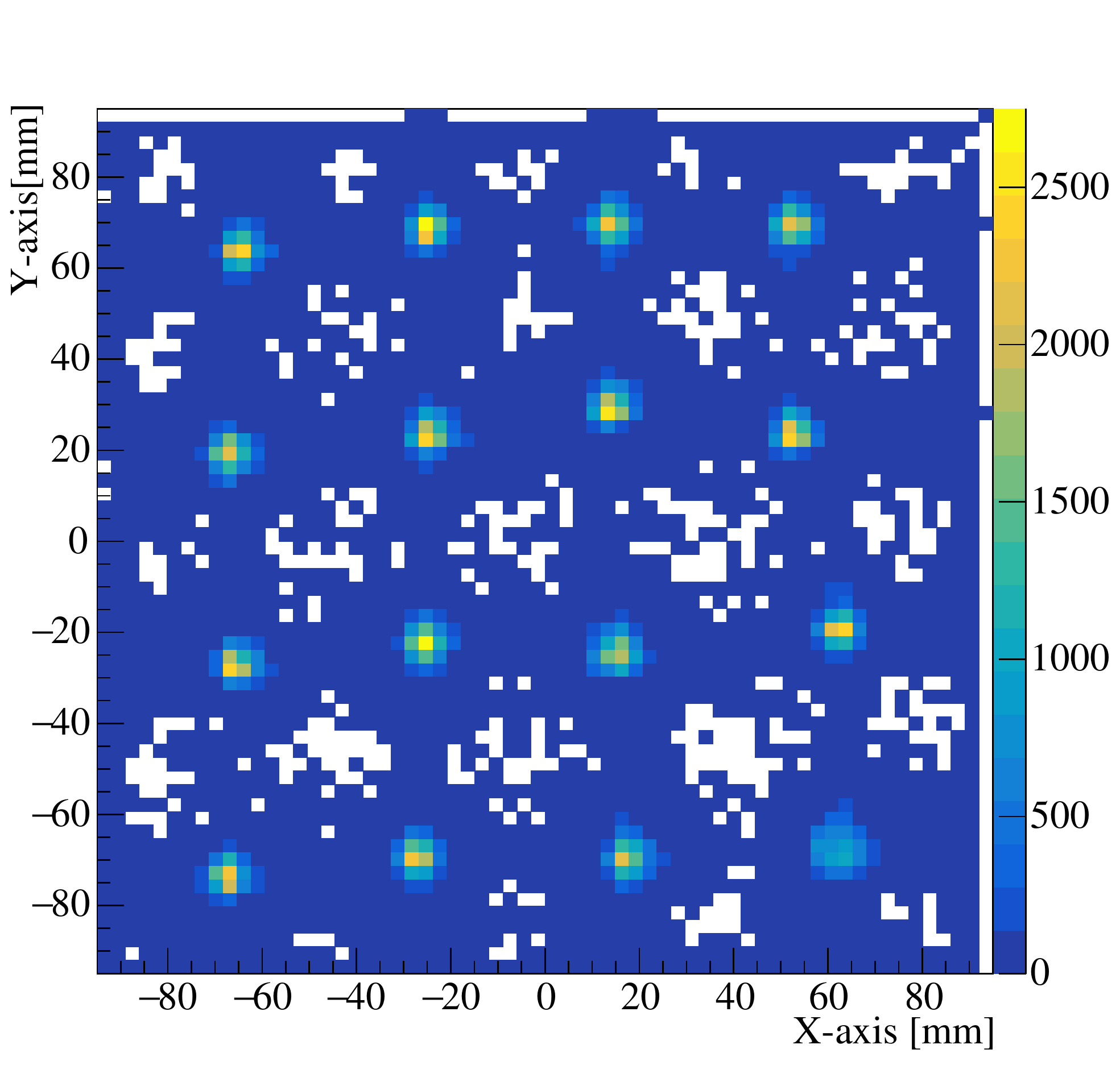}}
	\subfigure[Monte Carlo simulation]{
			\includegraphics[width=0.75\columnwidth]{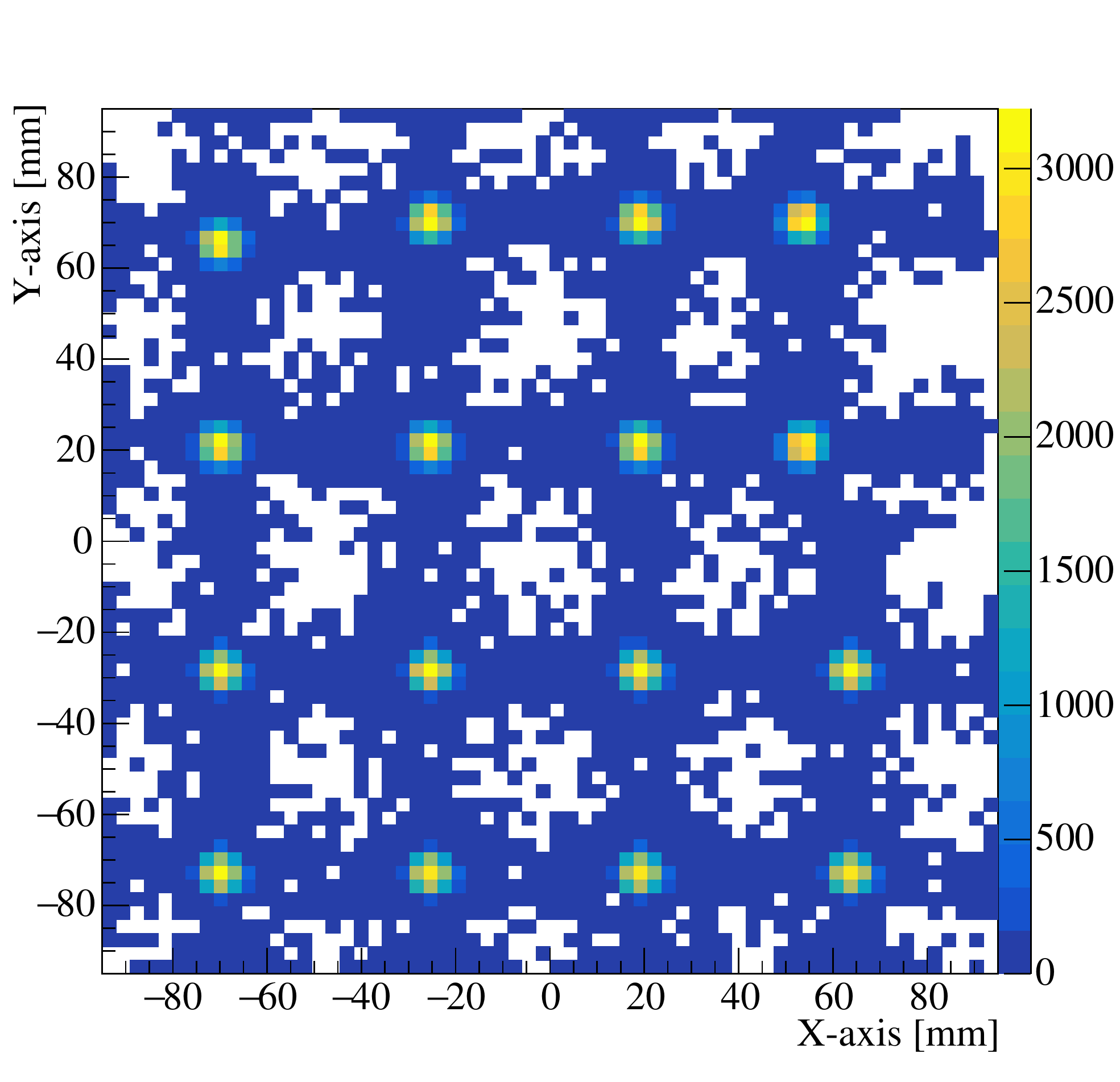}}
	\caption{Distribution of reconstructed starting points of alpha tracks from the $^{241}$Am source placed at 16 different locations on the cathode. 
	Fig. (a) is from experimental data while (b) is from Monte Carlo simulation data.	
	}
	\label{fig:mapping_originExSim}
\end{figure}

		
	
		
		

The $^{241}$Am source is used to study the distribution of the aforementioned parameters in our detector.
Fig.~\ref{fig:origin} shows the reconstructed starting points and the red square frame represents the physical boundary of the source, where the true starting points should fall within.
$74.3 \pm 0.9\%$ of the events have the starting point within the red frame, and $89.6\pm1.0\%$ within the yellow frame, which relaxes the boundary by one strip in each direction.
Only $1.6\pm0.1\%$ of the reconstructed starting points are 27~mm away from the center of the source (green square). 

Fig.~\ref{fig:source_angle} shows the distribution of the track orientation $\theta$ with respect to the vertical direction, in terms of $\cos(\theta)$.
While we expect all tracks from the source goes upward, $95.1\pm1.0\%$ of the events have the reconstructed $\cos(\theta)\gt0$.
In Fig.~\ref{fig:source_nHits}, the distribution of nHits for the tracks of $\alpha$ events is shown. 
$86.4\pm0.9\%$ of the events have nHits no smaller than 10.
When the three cuts are combined, we would select $81.9\pm2.0\%$ of the events from a calibration source with a small footprint in the TPC. 

Selection cut efficiencies based on tracks are also studied with Monte Carlo simulations based on the Geant4 toolkit~\cite{Geant4}.
The detector response to tracks is simulated with the REST framework~\cite{Altenmuller:2021slh} and details of our setup have been discussed in~\cite{Screener3D}.
For the $^{241}$Am source placed at one spot, the selection efficiency with three cuts applied is $85.0\pm1.5\%$. 

The cut efficiency dependence on position is simulated with the $^{241}$Am being placed at different locations on the cathode.
The active area of  $20\times20$~cm$^2$ is evenly divided into 16 regions and the $^{241}$Am source is placed at the center of each region for measurements.
Fig.~\ref{fig:mapping_originExSim} shows the reconstructed starting points of tracks.
The total track-related efficiencies from the experiment and simulation are  $65.5\pm3.2\%$ and $66.1\pm3.0\%$, respectively.
If we only consider the center four positions, the efficiencies are $79.3\pm1.6\%$ and $80.0\pm1.4\%$ for experiment and simulation, respectively.

If we consider a sample the same size as the cathode itself and U/Th radio-contamination is uniformly distributed on the surface of the sample, we can evaluate the track-related efficiency as $37.7\pm0.6\%$.

\section{Intrinsic TPC background}
\subsection{Background data taking}
\begin{figure}[tb]
	\centering
	\includegraphics[width=\columnwidth]{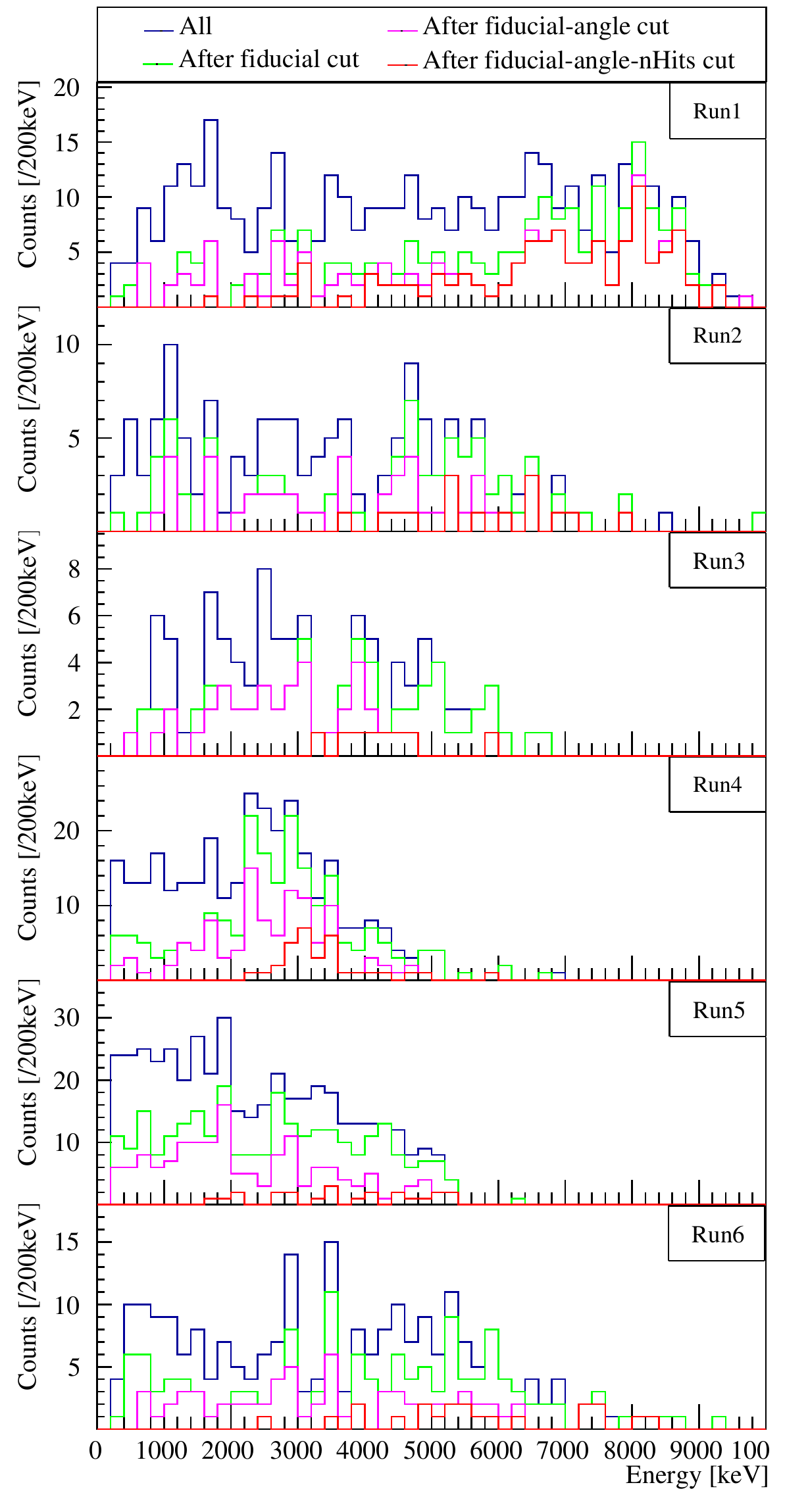}
	\caption{Energy Spectra of the background runs without and with track-related selection cuts.}
	\label{fig:EnergySpec}
	\end{figure}

\begin{table}[tb]
	\centering
	\caption{The run summary and count rates of the background runs. }
	\begin{tabular*}{\hsize}{@{}@{\extracolsep{\fill}}ccccc@{}}
		\hline
	
		Run    &flush rate &Run time    &Counts        &After Cuts\\
		       &L/min      &hour        &counts/hour   &counts/hour \\
		\hline  
		1     &0.55      &25.30         &16.56$\pm$0.81   &4.35$\pm$0.41\\		
		2     &0.55      &39.00         &3.87$\pm$0.32    &0.38$\pm$0.10\\      
		3     &0.35      &25.00         &4.20$\pm$0.41    &0.32$\pm$0.11\\   
		4     &0.73      &158.25        &2.06$\pm$0.11    &0.20$\pm$0.04 \\
		5     &0.20      &240.00        &1.88$\pm$0.09    &0.10$\pm$0.02  \\   
		6     &0.10       &90.00        &2.69$\pm$0.12    &0.23$\pm$0.03\\
		\hline
	\end{tabular*}
	\label{tab:RunInformationForTPC}
\end{table}

The intrinsic background of the detector is characterized by extended periods of physics data taking from February 28 to April 12, 2022.
During the entire data-taking period, the TPC takes data without any samples inside and is always kept intact even if data-taking is paused. 
We have observed the decreases in radon contribution over time and tested the impact of different flow rates on the background counting rates. 
Table~\ref{tab:RunInformationForTPC} summarizes the key information about the runs including the gas flow rate, data taking time, and count rates before and after tracking cuts.
Fig.~\ref{fig:EnergySpec} lists the energy spectra of all runs with and without the cuts.

During the assembly, the detector components are exposed to air, and radon contamination is inevitable. 
Once assembled, the detector is flushed with Ar-CO$_2$ mixture at a high rate of 3.5~L/min for 20 minutes before taking data.
Even when the data taking starts, we keep a relatively high flush rate of 0.55 L/min, especially considering the inner volume of the detector is only approximately 6~L.

\begin{figure}[tb]
	\centering
	\subfigure[The counting rate evolution of Run 1]
	{\includegraphics[width=0.75\columnwidth]{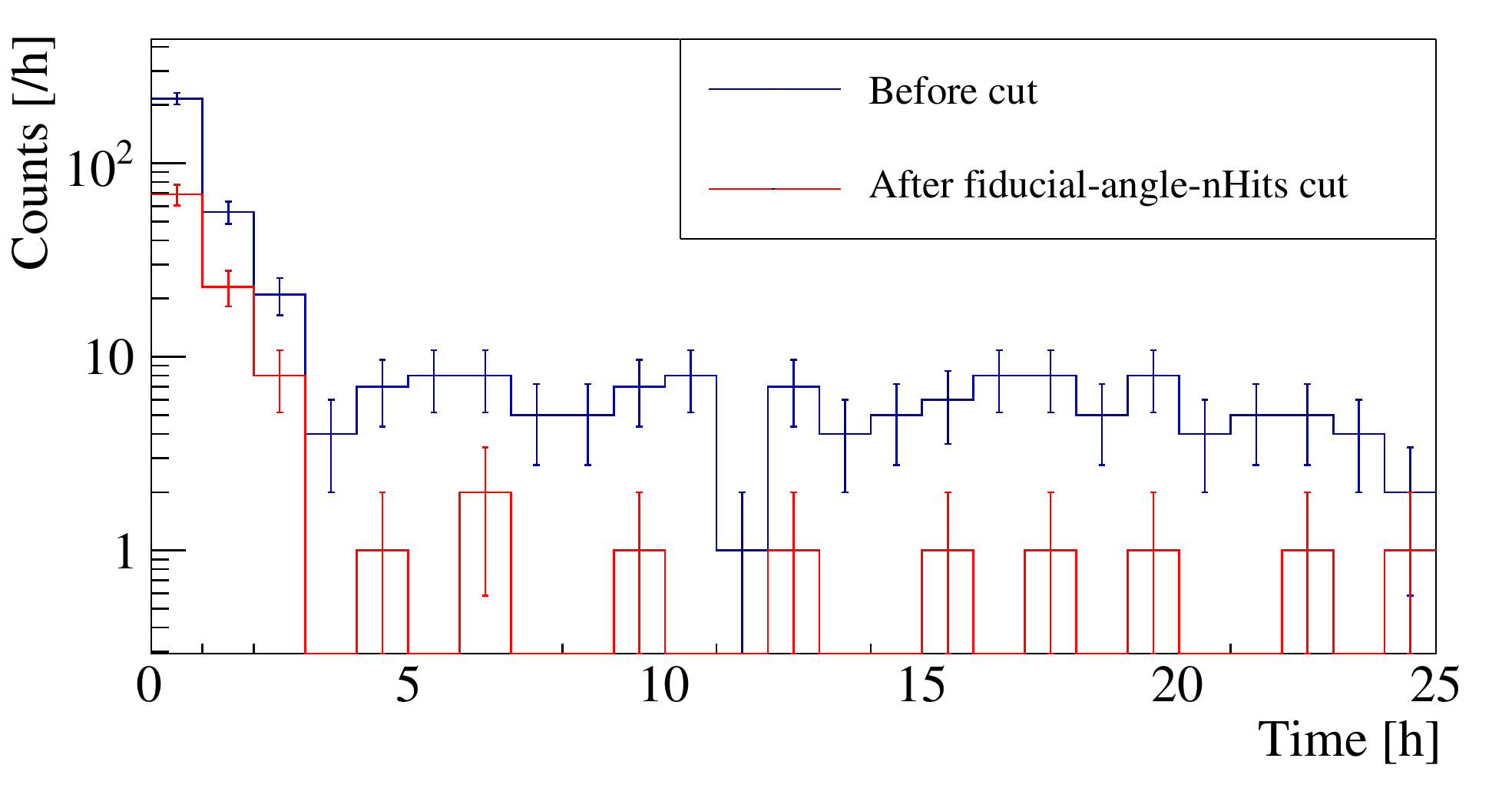}
		\label{fig:Run1_hour}
	}
	\subfigure[The rate in the first 3 hours of Run 1]
   {\includegraphics[width=0.75\columnwidth]{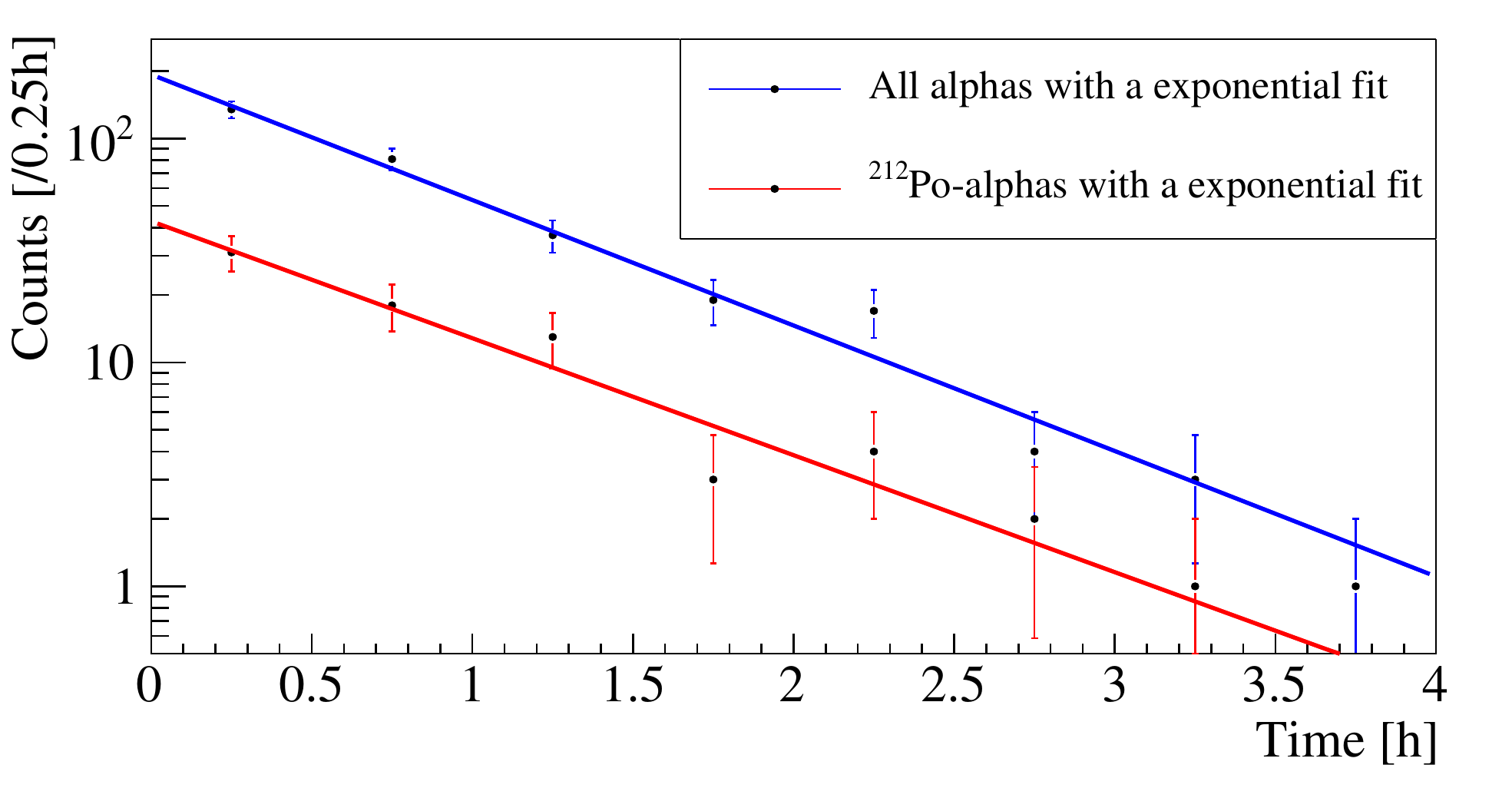}
	\label{fig:Run1_First3hour}
    }
	\subfigure[The counting rate evolution of Run 5]
	{\includegraphics[width=0.75\columnwidth]{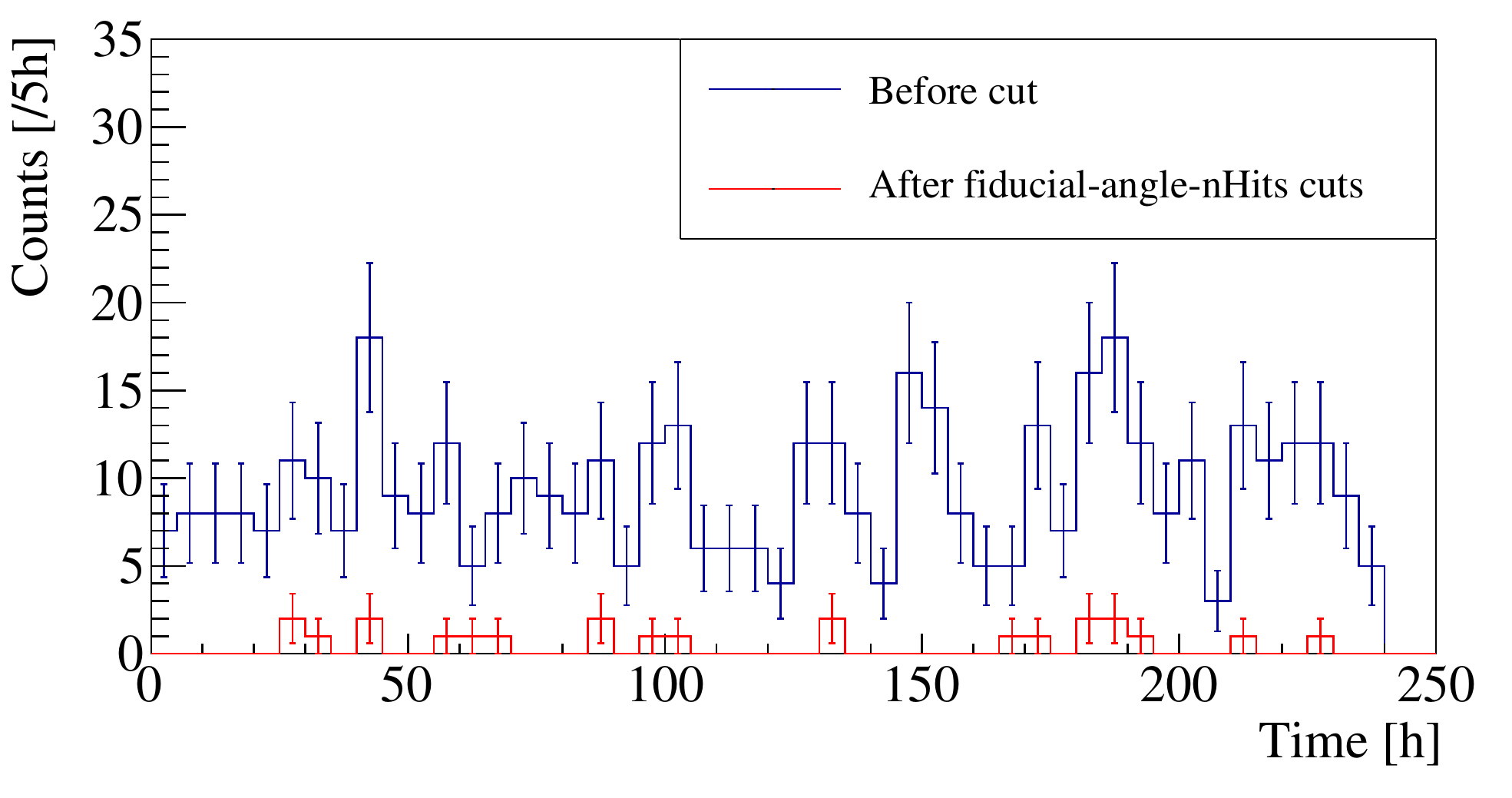}
		\label{fig:Run5_hour}}	
	\caption{The counting rate evolution with the time of Run 1 (a) and Run 5 (c). In Run 1, the counting rate drops rapidly in the first four hours as shown shown in (b). The solid lines in Fig. (b) represents exponential fits to data.}
	\end{figure}

Possible radon trapped inside the TPC is rapidly displaced from the TPC, as indicated by the rate decrease during the data taking.
The decrease is evident when comparing the event rate in Run 1 and later runs and comparing the rate evolution during the Run 1 data-taking period.
Fig.~\ref{fig:Run1_hour} shows the event rate over time during the data taking of Run 1 and a rapid decrease is observed within the first four hours of data taking start time. 
A zoomed-in plot shows the total and high energy (E$>$6~MeV) alpha event rates in Fig.~\ref{fig:Run1_First3hour}.
High energy alpha events are mostly from radon progenies $^{214}$Po and $^{212}$Po. 
The fitted exponential decay time constants are $32.3\pm2.0$ and $35.3\pm5.2$ minutes respectively.
The consistent alpha event rate evolution illustrates the process of radon disappearance inside the TPC.

Alpha event rates from Run 2 and Run 4 also show a downward trend but the characteristic time constant is much longer at $3.0\pm0.5$ days.
The fraction of high energy (E$>$6~MeV) alpha events also decreases from Run 2 to Run 4 (see Fig.~\ref{fig:EnergySpec}).
Both observations are consistent with the decay of $^{222}$Rn, which are attached to the surface of the detector components.

Alpha event rates in Run 5 reached a relatively stable state.
As Fig.~\ref{fig:Run5_hour} shows, the count rates with and without track-related cuts fluctuate around the average.
Furthermore, the high-energy alpha particle is noticeably absent in Run 5. 
The resulting $0.10\pm0.02$ counts/h can be explained by alphas from the decay chains of U/Th contaminations in the bulk of the cathode, where alpha particles lose part of the energy before escaping into the gas medium. 

In Run 6, the gas flush rate is lowered to 0.1~L/min and the alpha event rate picks up again.
More noticeably, the high energy alpha events re-appear, which can not be explained by the existing U/Th contaminations in the detector material. 
The flush rate is so low that air may creep back into the active volume of the TPC and re-introduces radon to the detector.
The assumption is later confirmed by detector performance characterization that is done after background data taking (see below).

The detector gain is measured again with the $^{241}$Am calibration source immediately after the background runs are concluded. 
With a gas flush rate of 0.2~L/min or more, the detector gain does not change and is within the range of gains measured before the background runs as shown in Fig.~\ref{fig:stability}.
Once we lower the gas flush rate to 0.1~L/min, a drop of $2.5\pm0.1\%$ for the gain is observed.
The phenomena are observed repeatedly and can be explained by gas quality deterioration due to air creeping in.

\subsection{The intrinsic background rate}
The intrinsic background rate of the detector is calculated with data in Run 5.
With the fiducial volume cut, track orientation cut, and the number of triggered strips cut, the final event rate recorded in the TPC is $0.10\pm0.02$~counts/h.
Considering the fiducial area of $14.6\times14.6$~cm$^2$, the background rate is $0.13\pm0.03~\mu$Bq/cm$^2$ ($(4.7\pm0.9)\times 10^{-4}$~counts/cm$^2$/h), slightly better than that quoted by the best-in-class commercial detector Ultralo-1800 ($5\times10^{-4}$~counts/cm$^2$/h)~\cite{UltraLo1800}.

If we consider the copper cathode as a sample and assume all the counts are from the cathode, the surface contamination rate of the sample is $0.33\pm0.07~\mu$Bq/cm$^2$ for the alpha particles.


\section{Summary and outlook}
Based on our proposal of a gaseous TPC with Micromegas readout for surface contamination assaying, we have constructed a prototype TPC with readily available components for alpha measurements.
The prototype TPC is equipped with a thermal bonding Micromegas with an active area of $20\times20$~cm$^2$.
The energy resolution we achieved with Ar-7\%CO$_2$ (Ar-2.5\%isobutane) gas mixture is 9.5\% (4.8\%) FWHM at 5485~keV with the correction of gain non-uniformity of the Micromegas. 
We have confirmed that the overall gain drift is 3.7\% during eight months, demonstrating that the detector can be used stably for surface contamination measurement over long periods.

Samples to be measured would be put on the cathode for high efficiency. 
The maximum dimension of the thin film sample is the same as the size of the Micromegas.
We have measured the counting efficiency to be higher than 99\% with two calibrated samples.
The efficiencies are cross-validated with a commercial Ortec silicon detector.
Features of alpha tracks are used to select signals and suppress backgrounds.
We have demonstrated that the combination of track starting point, track orientation, and the number of triggered strips can effectively identify events originating from the samples at over 30\% efficiency.

We have commissioned the detector and measured the intrinsic background over 40 days of running. 
A significant decrease in background rates is observed since the detector starts running, due to displacement of residual radon in the detector and/or radon emanated from detector components. 
The detector reaches a stable background rate if the gas flush rate is maintained at 0.2~L/min or higher. 
The background rates are further reduced by 19 times, with the help of track-related cuts, to reach a background level of $0.13\pm0.03~\mu$Bq/cm$^2$.
If we assume all the background is from the copper cathode itself, we have measured the surface contamination rate of the cathode to be $0.33\pm0.07~\mu$Bq/cm$^2$ for the alpha particles. 

The prototype detector demonstrates the advantage of the Screener3D proposal based on TPC technology with Micromegas.
The combination of a large active area and tracking capability offers significant benefits in terms of intrinsic detector background and screening sensitivity.
The current background rate is better than the leading commercial solutions.
A future iteration of the Screener3D detector will have 6 Micromegas modules tiled together to improve the effective screening area by an order of magnitude.
A better energy resolution is also expected with an improved version of the thermal bonding Micromegas and better choices of working gas mixture. 
Detector components would be selected with an emphasis on bulk and surface contamination levels to further reduce the background rate of the detector. 
We will also optimize the operational procedure to facilitate the screening workflow.
The Screener3D detector will be used for surface contamination screening for underground experiments for neutrinoless double beta decay, dark matter, and other rare event searches.   

\section*{Acknowledgements}
This work was supported by the grant from the Ministry of Science and Technology of China (No.~2016YFA0400302) and the grant U1965201 from the National Natural Sciences Foundation of China. We appreciate the support from the Chinese Academy of Sciences Center for Excellence in Particle Physics (CCEPP).

\bibliography{sn-bibliography}

\end{document}